\documentclass[12pt,a4paper]{panl}
\usepackage{cite}
\usepackage{wrapfig}
\usepackage{graphicx}
\usepackage{amssymb}
\usepackage{amsfonts}
\usepackage{amsmath}
\usepackage{longtable}
\usepackage{rotating}
\usepackage{lscape}
\usepackage{multirow}

\originalTeX
\begin{document}

\title{Оpen and hidden strangeness with kaons and $\varphi$-mesons in Bjorken energy density approach for central A+A collisions from SPS to LHC}
\maketitle
\authors{O.\,Shaposhnikova$^{a,b}$\footnote{E-mail: shaposhnikova.om23@physics.msu.ru},
A.\,Marova$^{b}$\footnote{E-mail: st097602@student.spbu.ru},
G.\,Feofilov$^{b}$\footnote{E-mail: g.feofilov@spbu.ru}}
\setcounter{footnote}{0}
\authors{О.М.\,Шапошникова$^{a,b}
$\footnote{E-mail: shaposhnikova.om23@physics.msu.ru(русский вариант)},
А.А.\,Марова$^{b,2}
$\footnote{E-mail: st097602@student.spbu.ru(русский вариант)},
Г.А.\,Феофилов$^{b,2}
$\footnote{E-mail: g.feofilov@spbu.ru(русский вариант)}}
\from{$^{a}$\,Moscow State University}
\from{$^{a}$\,Московский государственный университет}
\from{$^{b}$\,Saint-Petersburg State University}
\from{$^{b}$\, Санкт-Петербургский государственный
 университет, Россия, 199034, Санкт-Петербург, Университетская наб. 7/9}

\begin{abstract}

 С целью сравнения  вкладов в плотность энергии Бьоркена мы используем имеющиеся данные о значениях $<dN/dy>$ и $<\it{p}_T>$  для адронов, в том числе для  \(\pi^{+}\)+ \(\pi^{-}\) , \(K^{+}\)+ \(K^{-}\), \(p\)+ \(\overline{p}\), $K^*(892)^0$ и $\varphi$-мезонов, зарегистрированных в области нулевых быстрот $(\vert y\vert < 0.5)$ в центральных 0-5\% столкновениях Au-Au, Pb-Pb и Xe+Xe в широком диапазоне энергий. Частицы типа странно-нейтрального $\varphi$-мезона (система \(s\overline{s}\) кварков) и K-мезона (содержащего одиночный s-кварк) представляют особый интерес, поскольку они могут иметь разные  механизмы  рождения и чувствительности к свойствам кварк-глюонной плазмы.
 \
\vspace{0.2cm}

 We use the available data  on $<dN/dy>$ and $<\it{p}_T>$  for  the identified hadrons including  \(\pi^{+}\)+ \(\pi^{-}\), \(K^{+}\)+ \(K^{-}\), \(p\)+ \(\overline{p}\), $K^*(892)^0$ and $\varphi$-mesons, registered  at midrapidity  $(\vert y\vert < 0.5)$ in  central 0-5\% Au-Au, Pb-Pb and Xe+Xe collisions in a broad range of energies in order to compare the relative  contributions to the Bjorken energy density. Particles, like strangeness-neutral $\varphi$-meson (a system of \(s\overline{s}\) quarks) and K-meson (containing single s-quark),   are of specific interest because they might have different production mechanisms and differ in sensitivity to the properties of the QGP-medium  formed in relativistic heavy-ion collisions.

\end{abstract}
\vspace*{6pt}


\label{sec:intro}
\section*{Introduction}

 According to the prediction in  \cite{ PhysRevLett.48.1066} by J.Rafelski and B. M\"uller,      the formation of quark-gluon plasma (QGP) in relativistic nucleus-nucleus collisions would result in an increased yield of particles containing strange quarks, if compared to the proton-proton case.  The formation of $s$ and $\overline{s}$ quarks can be  increased in the QGP due to the multiple perurbative gluon-gluon interactions of the $gg\rightarrow$ $s\overline{s}$ type, as well as in the processes \(u\overline{u}\), \(d\overline{d}\) → \(s+\overline{s}\). Non perturbative processes could dominate the production of strange hadrons at low $p_T$ where the phenomenogical models with fragmentation of quark-gluon strings could be applied with the account of some  collectivity effects.
 
 The decisive role of open strangeness as a characteristic feature and signal of the formation of quark-gluon plasma was confirmed, in particular, in experiments such as NA57 \cite{Antinori_2006} and NA51/SHINE  \cite{na61collaboration2007na61shine, refId0,
 ADUSZKIEWICZ201735}  at SPS at CERN and in STAR \cite{Caines_2004}, \cite{Shi_2017} at the collider RHIC at BNL.  An increased yield of strangeness was also found later by ALICE at the LHC  in the high multiplicity events  collisions  of small systems ($pp$ and $p+Pb$) \cite{Adam-2017}.
 
The short-lived $\ensuremath{\varphi}$-meson is of particular interest since it is the lightest of the vector mesons with the  hidden flavor. The quark composition of $\varphi$-meson  can be thought of as a mixture of \(s\overline{s}\), \(u\overline{u}\) and \(d\overline{d}\) states, but it is  considered to be very close to the pure \(s\overline{s}\) state. 

The OZI rule \cite{PhysRevD.16.2336} strongly influences the rate of birth and decay of $\ensuremath{\varphi}$-meson, suppressing its decay to three pions and causing it to predominantly decay in a pair of \(K^{+}\)+ \(K^{-}\) mesons.

 Additionally, \cite{PhysRevLett.54.1122} suggests that the absence of OZI suppression can lead to a significant increase in the production of $\varphi$-mesons during the formation of the QGP. You can find a review of measurements of $\varphi$-meson yields in \cite{Nasim-2015}.

We present below a comparative analysis and discuss the $\sqrt{{s_{NN}}}$ dependences of the contributions by the identified  particles like $\ensuremath{\varphi}$-mesons and   $K^*(892)^0$  resonances and, as well as by \(K^{+}\)+ \(K^{-}\),  \(\pi^{+}\)+ \(\pi^{-}\) and  \(p\)+ \(\overline{p}\) hadrons  to the  Bjorken energy density related to the interaction region in the most central ( 0-5\%) nucleus-nucleus collisions.

\label{sec:bjorken}
\section{ Bjorken energy density}

Following \cite{PhysRevD.27.140}, the Bjorken energy density $\epsilon$  (\ref{eq1}) is determined at midrapidity through the mean transverse energy  $dE_{\perp }/{dy}$ by the particles   formed in the volume of a cylinder with a cross-sectional area $S_{\perp }$ determined by the overlap between colliding nuclei and the length corresponding to the characteristic particle formation time $\tau$:

\begin{equation}\label{eq1} 
  \epsilon=\frac{dE_{\perp }}{dy}\cdot\frac{1}{S_{\perp }\tau}
\end{equation}

 With the transverse mass of an identified hadron $\langle m_{\perp}\rangle=\sqrt{m^2+{p_{T}}^2}$, one can approximate the mean transverse-energy rapidity density, relevant to the given centrality class of collisions, with 
a sum (\ref{eq2}) of contributions by different  hadrons. 
 In our estimates of components of ${\frac{dE_{\perp }} {dy}}$, we used the available published data with the experimental values of $<dN/dy>$ and $<\it{p}_{T}>$ obtained in 
\cite{PhysRevC.78.044907,PhysRevLett.87.052301,PhysRevC.88.044910,2023137730, PhysRevC.94.034903,PhysRevC.101.044907}.  This includes also resonances $K^*(892)^0$ and $\ensuremath{\varphi}$-mesons registered in the central rapidity region 
in a wide energy range of nucleus-nucleus collisions (from the SPS to  the LHC). 

The majority of Bjorken energy density estimates are widely using the quantity of $\epsilon\cdot\tau$ (\ref{eq3}) with the particle formation time $\tau=1$ fm/c. Our study also follows  this approach. 

We have to note also that the  Bjorken energy density approach may not be entirely accurate at the energies below RHIC, because the formation time ($\tau$) should be larger than the time of passage of the colliding nuclei through each other \cite{Litvinenko-2007}.

\begin{equation}\label{eq2} 
  {\frac{dE_{\perp }} {dy}}=\frac{3}{2}(\langle m_{\perp}\rangle\frac{dN}{dy})_{\pi_-^+}+2(\langle m_{\perp}\rangle\frac{dN}{dy})_{K_-^+, p, \bar{p}}+(\langle m_{\perp}\rangle\frac{dN}{dy})_{K^*(892)^0}+(\langle m_{\perp}\rangle\frac{dN}{dy})_{\varphi}
\end{equation}

\begin{equation}\label{eq3} 
  \epsilon\cdot\tau=\frac{dE_{\perp }}{dy}\cdot\frac{1}{S_{\perp }}
\end{equation}

  The quantity of $\epsilon\cdot\tau$ (\ref{eq3}) still contains another parameter - the transverse area $S_{\perp }$, one cannot measure it  directly . The usual methods to define it are based on the Glauber Monte Carlo (GMC) approach combined with the multiplicity distribution analysis  where  the  estimated mean number of participating nucleons   ($N_{part}$) is considered  to be in the relation to the transverse area \cite{na61collaboration2007na61shine, Adam-2017}. 
  
 In our study, instead of the GMC,  we assume that the measured particle production is straightforwardly related  to the   geometry of partially overlapping nuclei.

 Calculations, assuming colliding disks of radii R = 6.87 fm \cite{Caines_2004} for Au–Au collisions (or $R = 7.17$ fm \cite{PhysRevC.94.034903} for Pb–Pb collisions),  show that in both cases the dominant events selected in a given central 0–5\% class will have a noticeable shift of  the average impact parameter $<b>$  from 0. The following values  could be obtained for the class 0–5\% Au–Au collisions: $< b > = 2.18$ fm,  and for  0–5\%  Pb–Pb collisions -- $< b > = 2.26$ fm.  (In both cases we estimate the accuracy of $< b >$ value to be $\sim0.2$ fm). Naturally, this  will lead to a smaller value of the overlap area  if compared to $< b > = 0$. The    GMC calculations  for central Au-Au collisions that gave  also similar  results for $< b >$ $\sim $ 2.2, 2.39, 2.21 fm at three RHIC energies,  see \cite{Caines_2004}.  Therefore,  in our approach we use the following corrected  value of the mean initial transverse area $S_{\perp }$  for the given classes of central events 0–5\% of Au-Au and Pb-Pb collisions:  $118.5 \pm 2.9 $ $fm^2$ and $129.5 \pm 2.2$ $fm^2$.  Naturally, this results in about 20\%  higher values of $\epsilon\cdot\tau$ then in the usually applied GMC. (We may also argue that our  straightforward aproach should gave a smaller systematic bias on the final results).
 
In case of collisions of the deformed Xe nuclei $(\beta = 0.18\pm0.02)$ at 5.44~TeV, we used  the efficient value $S_{{eff}{\perp }}$. We obtained it  after  the normalization of the  relevant values of $\epsilon\cdot\tau$ on the data for  pions extrapolated to the energy $\sqrt{s_{NN}}=5.44$ TeV. This gives for the overlap area of for  0–5\% centrality of  Xe-Xe collisions the value   $S_{{eff}{\perp }} = 70.1 \pm$ 15.6 $fm^2$,   that is used further  to calculate  the contributions to $\epsilon\cdot\tau$ by  \(K^{+}\)+ \(K^{-}\), \(p\)+ \(\overline{p}\), $K^*(892)^0$ and $\varphi$-mesons.

\label{sec:results}
\section*{Results}

We show in Fig. \ref{fig01} the results of energy dependence of fractions of   $\epsilon\cdot\tau$ as defined in  (\ref{eq4}) and  obtained in our study for the identified hadrons including  \(\pi^{+}\)+ \(\pi^{-}\), \(K^{+}\)+ \(K^{-}\), \(p\)+ \(\overline{p}\), $K^*(892)^0$ and $\varphi$-mesons. 
The identified hadrons were measured at midrapidity  in  central 0-5\% Au-Au, Pb-Pb and Xe+Xe collisions in a broad range of collision energies.

\begin{figure}[h]
\begin{center}
\includegraphics[width=127mm]{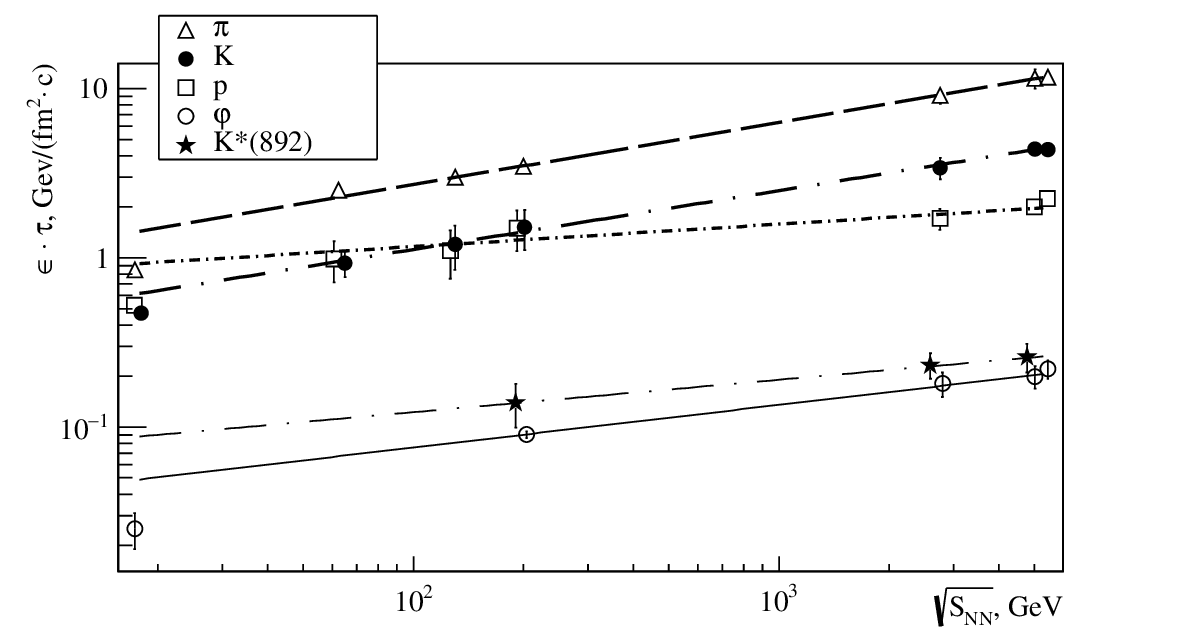}
\vspace{-3mm}
\caption{Values of Bjorken  $\epsilon\cdot\tau$ with $\tau=1$ fm/c in  {$0-5\%$} Au-Au, Pb-Pb and Xe-Xe collisions as a  function of $(\sqrt{s_{NN}})$ for identifed hadrons  (including $\pi^{+}+\pi^{-}$, $K^{+}+K^{-}$, $p+\overline{p}$, $K^*(892)^0$ and $\ensuremath{\varphi}$-meson).  Estimates were done  using published data  on $<p_t>$ and $<dn/dy>$  measured in \cite{PhysRevC.78.044907,PhysRevLett.87.052301,PhysRevC.88.044910,2023137730,PhysRevC.94.034903,PhysRevC.101.044907} at midrapidity ($(\vert y\vert < 0.5)$, except data at 17.3 Gev, where $(\vert y\vert < 0.6)$). Lines -- are the results of power-law fits (\ref{eq4}). We show only statistical uncertainites. We estimate total systematic uncertainties  to be $\le 10\%$ for Au-Au and Pb-Pb data and  $\le 25\%$ in case  of Xe-Xe. }
\end{center}
\labelf{fig01}
\vspace{-5mm}
\end{figure}

Power-law fits (\ref{eq4})  were  performed  for $\epsilon\cdot\tau$ vs. $\sqrt{s_{NN}}$  data in the region 
$\sqrt{s_{NN}}$ ~  of 200 GeV -- 5.2 TeV. (We note that we take in (\ref{eq4}) the quantity $s$  as dimensionless $s/ s_0$, where $s_0=1$ $GeV^2$). One may find the fit parameter values in Table \ref{Table1}. 
 
\begin{equation}\label{eq4} 
  \epsilon\cdot\tau=Q\cdot (s_{NN})^n
\end{equation}

We observe in Fig.1:
\begin{itemize}
\item { in case of pions, the $s$-dependence of  $\epsilon\cdot\tau$ is    growing  faster ($s^{0.184\pm0.015}$) then the power-law behaviour in  case of the  charged particle multiplicities in  AA collisions well described by the function $s^{0.155}$ \cite{PhysRevLett.116.222302}};

\item {for pions, kaons and protons, the $s$-dependences of  $\epsilon\cdot\tau$ demonstrate different behavior from one to another. This is also true   for $\ensuremath{\varphi}$-mesons  ($s^{0.126\pm0.020}$)  in comparison to $K$-mesons ($s^{0.17\pm0.03}$});

\item {the $s$-dependences of  $\epsilon\cdot\tau$ for two resonances -- $\ensuremath{\varphi}$-mesons and $K^*(892)^0$,  are  remarkably similar in the whole energy range (see Fig.\ref{fig01} and Table~\ref{Table1});} 

\item {one may note here that the masses of resonances -- $\ensuremath{\varphi}$-mesons and $K^*(892)^0$ -- (1020 MeV  and 892 MeV)  are  rather  close to each other, thus we have a similar behaviour of $\epsilon\cdot\tau$ with energy. However,  the mass or proton is even closer to the mass of $\ensuremath{\varphi}$-meson, while the  contribution to the $s$-dependence of  $\epsilon\cdot\tau$ by protons is noticeably higher then for $\ensuremath{\varphi}$-mesons or $K^*(892)^0$.}
\end{itemize}


\begin{table}[ht]
\caption{Parameters of power-law $Q\cdot (s)^n$ approximatioms   for values of    $\epsilon\cdot\tau$ vs. $\sqrt{s}$ for (\(\pi^{+}\)+ \(\pi^{-}\)), (\(K^{+}\)+ \(K^{-}\)), (\(p\)+ \(\overline{p}\)), $K^*(892)^0$ and $\varphi$-mesons.}
\centering
  \begin{tabular}{ |p{3cm}|p{3cm}|p{3cm}|p{3cm}|  }
 \hline

  & n  & Q &$\chi^2/NDF$\\
 \hline
 $\pi$   & $ 0.184\pm0.015$    &$0.50\pm 0.09$  &$0.013/2$\\
 K &   $0.17\pm0.03$  & $0.23\pm0.11$   &$0.14/2$\\
 p &$0.07\pm0.03$ & $0.6\pm0.3$&  $0.6/2$\\
 $K^*(892)^0$ &   $0.10\pm0.05$  & $0.05\pm0.04$ &$0.001/1$\\
 $\varphi$   &$0.126\pm0.020$ & $0.024\pm0.006$&  $0.05/1$\\
 \hline
\end{tabular}
  \label{Table1}
\end{table}


\begin{table}[h]
\caption{Parameters of linear approximatioms $A+B\cdot (s)$  of ratios  of yields of  $\varphi$-mesons to  the relevant yields of (\(\pi^{+}\)+ \(\pi^{-}\)), (\(K^{+}\)+ \(K^{-}\)), (\(p\)+ \(\overline{p}\)) and  $K^*(892)^0$.}
\centering
  \begin{tabular}{ |p{1.3cm}|p{2.63cm}|p{2.5cm}|p{2.5cm}|p{2.6cm}| }
 \hline

    &
 $(\epsilon\cdot\tau)_{\varphi}/(\epsilon\cdot\tau)_{\pi}$ & $(\epsilon\cdot\tau)_{\varphi}/(\epsilon\cdot\tau)_{K}$  & 
   $(\epsilon\cdot\tau)_{\varphi}/(\epsilon\cdot\tau)_{p}$  & $(\epsilon\cdot\tau)_{\varphi}/(\epsilon\cdot\tau)_{K^*}$  \\
 \hline
 A &$0.026\pm0.002$ & $0.06\pm0.01$& $0.05\pm0.01$ & $0.7\pm0.2$\\
 B,$10^{-6}$ &$-1.5\pm0.5$ & $-1.6\pm2.1$ & $9.2\pm2.7$ & $26.0\pm55.0$ \\
 $\chi^2/ndf$ &$0.8/3$ & $0.4/3$ & $1.8/3$ & $0.11/1$ \\

 \hline
\end{tabular}
  \label{Table2}
\end{table}


In case of the hidden stranges formation, we see in Fig.\ref{fig01} and with   Table~\ref{Table1}  data that  the Bjorken energy fraction of $\epsilon\cdot\tau$, relevant to $\varphi$-mesons,  is growing much slower  with the  collision energy ($s^{0.126\pm0.020}$) then the one relavant to pions ($s^{0.184\pm0.015}$). We can qualitatively explaine this for pions as a result of growth with the collision energy of the mean  $p_{\perp}$, that contributes to the transverse mass.  As to the  case of \(p\)+\(\overline{p}\), we see  rather slower growth ($s^{0.17\pm0.03}$)   with $\sqrt{s_{NN}}$.  Taking into account the fact that  masses of  $\ensuremath{\varphi}$-mesons and $K^*(892)^0$ -- (1020 MeV  and 892 MeV)  are close to the mass of   proton (938 MeV) one may asume that the partonic degrees of freedom in QGP are playing a different role in the  formation of protons and resonances.  

Another observation concerns the ratios of  fractions of Bjorken energy densitity $\epsilon\cdot\tau$  values as a function of collision energies obtained in this work for several identified hadrons. (It is evident that these are the ratios of the relevant transverse energies). In Fig.\ref{fig02} we show the ratios of values of  $\epsilon\cdot\tau$   of $\varphi$-mesons to the relevant densities for other hadrons: (\(\pi^{+}\)+ \(\pi^{-}\)), (\(K^{+}\)+ \(K^{-}\)),  (\(p\)+ \(\overline{p}\)) and  $K^*(892)^0$. 

In Table~\ref{Table2} we present the  paraemetrs of approximation of ratios of  particle yields with linear function  $A+B\cdot (s)$ .

As we see, in the whole region of collision energies under study, these ratios  of  particle yields $(\epsilon\cdot\tau)_\varphi /({{{(\epsilon\cdot\tau)_{(\pi^{+}+\pi^{-}),(K^{+}+K^{-}),(p+\overline{p},K^*(892)^0)}}}})$ are   almost  flat with $\sqrt{s_{NN}}$ (see Fig.{\ref{fig02}}). A similar flat dependence, but {\it {for the ratios of the  yields}} of $\varphi$-mesons to  yields of $K^{-}$

was observed earlier in the domain from SPS to RHIC energies (see the compilation of data and analysis in \cite{Bratkovskaya-2023}) and at the LHC \cite{PhysRevC.106.034907}. Our results of the ratio of the  yields 
$2\cdot\varphi/(K^{+}+K^{-}) =0.12\pm0.2$ is  very near to the mean ratio of yields  $\varphi/K^{-}=0.15\pm0.3$ observed in a wide region of collision energies  (see in  \cite{Bratkovskaya-2023} and  \cite{PhysRevC.106.034907}).

Among the theoretical models directly focused on medium effects on hidden strangeness production in heavy-ion collisions, we mention just briefly  the microscopic Parton-Hadron-String Dynamics (PHSD) transport approach  model          \cite{ Bratkovskaya-2023} that was successfully applied  to describe the ratios  of $\varphi$-mesons to  yields of $K^{-}$ observed earlier in the domain from SPS to RHIC energies. PHSD  model is a non-equilibrium microscopic transport approach for the description of the dynamics of strongly interacting hadronic and partonic matter. At the region of the LHC  energies, the ratio of yields  $\varphi$-mesons to  yields of $K^{-}$ was  described in the statistical hadronization model \cite{Stachel_2014}. One may find  also in  \cite{PhysRevC.106.034907} discussion of a  number of different approaches  like  EPOS3 model calculations with and without a hadronic cascade phase modeled by UrQMD. 
We may conclude here that the complete understanding of a hidden strangeness production  and of the medium produced effects of strangeness  still are not quite well understood and this requires the additional studies with new observables. 

\begin{figure}[t]
\begin{center}
\includegraphics[width=127mm]{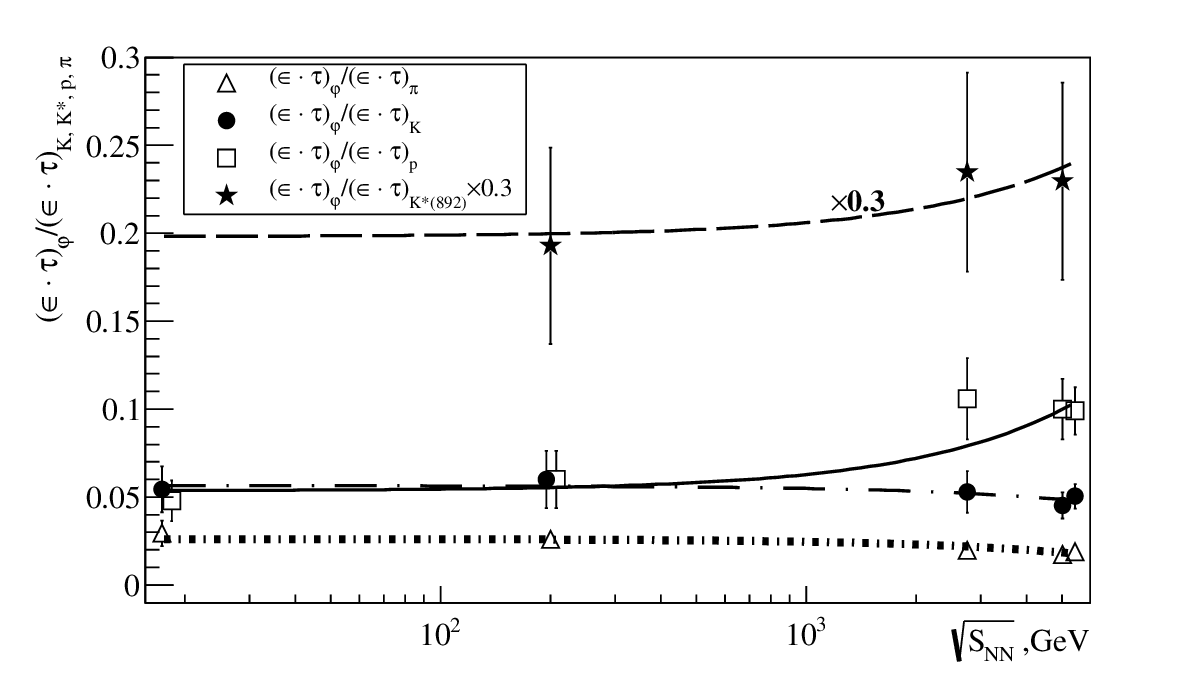}
\vspace{-3mm}
\caption{  A) Energy dependence for ratios of values of Bjorken energy densities  
$(\epsilon\cdot\tau)_\varphi/(\epsilon\cdot\tau)_{(\pi^{+}+\pi^{-}),(K^{+}+K^{-}),(p+\overline{p},K^*(892)^0)}$.
Lines are linear fits  ${a\cdot{s_{NN}}+b} $ (see Table~\ref{Table2}).
}
\end{center}
\labelf{fig02}
\vspace{-5mm}
\end{figure}

\hfill \break
\section*{ Conclusions}
We calculated the values of mean Bjorken energy density $\epsilon\cdot\tau$  using the available published data with experimental values of $<dN/dy>$ and $<\it{p}_T>$ obtained in  (\cite{PhysRevC.78.044907} -- \cite{PhysRevC.101.044907} ) for  several identified hadrons including  strangeness-- neutral  $\varphi$-meson (\(s\overline{s}\) quark system), $K^*(892)^0$ and  \(K^{+}\)+ \(K^{-}\)  mesons (containing single s (or $\overline{s}$)-quark). We  obtained  estimates for central 0–5\% classes of  Au–Au, Pb–Pb and Xe-Xe collisions in a wide  range of $\sqrt{s_{NN}}$ -- up to the highest at the LHC. We took also into account in our calculations of $\epsilon\cdot\tau$  for all colliding systems the fact that  in any of  central  0–5\% class the mean impact parameter value  is shifted from 0 to the value of $\sim2.3\pm0.2$ fm. Therefore, we corrected accordingly the transverse area of the interaction region $S_\perp$ for central 0–5\% classes for all colliding systems used in this analysis. 

 We observe different  dependencies on the collision energy for the fractions of  the Bjorken energy density $\epsilon\cdot\tau$. We found that the one relevant to  $(\pi^{+}+\pi^{-})$  grows with $s$ faster  ($s^{0.184\pm0.015}$)  then in case of multiplicity ($s^{0.157}$) \cite{PhysRevLett.116.222302}. This could be explained  by the contribution from the  well known growth  in AA collisions of the $<p_T>$ value for pions  with $\sqrt{s_{NN}}$. At the same time, the results for $(p+\overline{p})$ demonstrate the slowest dependence on the energy of collisions: ($s^{0.07\pm0.3}$).
 
 We observe also that the ratios of the Bjorkrn energy fractions, relevant to the identified hadrons -  $(\epsilon\cdot\tau)_\varphi /({{{(\epsilon\cdot\tau)_{(\pi^{+}+\pi^{-}),(K^{+}+K^{-}), K^*(892)^0)}}})}$, in all cases, are practically not depending on the collision energy. This behavior looks similar to the one previodly observed in  {\it{the ratios of  yields}} of $\varphi$-mesons  to $K^{-}$ mesons.

It is in our plans  to include in these studies  of Bjorken energy densities the additional available data  of  $\lambda$ hyperons, as well as of the  particles containing two or three strange quarks, to compare with the hidden strangeness production in relativistic heavy-ion collisions.

 Acknowledgement:
 Supported by Saint Petersburg State University, project ID: 94031112.
 
 Authors are grateful to Vladimir Kovalenko for useful discussions.

\bibliographystyle{pepan}
\bibliography{pepan_biblio}

\begin{thebibliography}{10}
\def\selectlanguageifdefined#1{
\expandafter\ifx\csname date#1\endcsname\relax
\else\selectlanguage{#1}\fi}
\providecommand*{\href}[2]{{\small #2}}
\providecommand*{\url}[1]{{\small #1}}
\providecommand*{\BibUrl}[1]{\url{#1}}
\providecommand{\BibAnnote}[1]{}
\providecommand*{\BibEmph}[1]{\emph{#1}}
\ProvideTextCommandDefault{\cyrdash}{\hbox to.8em{--\hss--}}
\providecommand*{\BibDash}{}

\bibitem{PhysRevLett.48.1066}
\selectlanguageifdefined{english}
\BibEmph{Rafelski J., M\"uller B.} Strangeness Production in the Quark-Gluon Plasma~// \href{http://dx.doi.org/10.1103/PhysRevLett.48.1066}{Phys. Rev. Lett.} \BibDash
\newblock 1982. \BibDash Apr. \BibDash
\newblock V.~48. \BibDash
\newblock P.~1066--1069. \BibDash
\newblock URL: \BibUrl{https://link.aps.org/doi/10.1103/PhysRevLett.48.1066}.

\bibitem{Antinori_2006}
\selectlanguageifdefined{english}
\BibEmph{Antinori F., Bacon P., A~Badal$\grave{a}$~et al. N.C.} Enhancement of hyperon production at central rapidity in 158 A GeV/c PbPb collisions~// \href{http://dx.doi.org/10.1088/0954-3899/32/4/003}{Journal of Physics G: Nuclear and Particle Physics}. \BibDash
\newblock 2006. \BibDash feb. \BibDash
\newblock V.~32, no.~4. \BibDash
\newblock P.~427. \BibDash
\newblock URL: \BibUrl{https://dx.doi.org/10.1088/0954-3899/32/4/003}.

\bibitem{na61collaboration2007na61shine}
\selectlanguageifdefined{english}
\BibEmph{Laszlo A. et~al.} [NA61 Collaboration] NA61/Shine at the CERN SPS. \BibDash
\newblock 2007. \BibDash
\newblock arXiv:0709.1867~[nucl-ex].

\bibitem{refId0}
\selectlanguageifdefined{english}
\BibEmph{{Balkova, Yuliia}.} Strangeness production in the NA61/SHINE experiment at the CERN SPS energy range~// \href{http://dx.doi.org/10.1051/epjconf/202227102013}{EPJ Web Conf.} \BibDash
\newblock 2022. \BibDash
\newblock V. 271. \BibDash
\newblock P.~02013. \BibDash
\newblock URL: \BibUrl{https://doi.org/10.1051/epjconf/202227102013}.

\bibitem{ADUSZKIEWICZ201735}
\selectlanguageifdefined{english}
\BibEmph{Aduszkiewicz A. et~al.} [NA61 Collaboration] Recent results from NA61/SHINE~// \href{http://dx.doi.org/https://doi.org/10.1016/j.nuclphysa.2017.04.046}{Nuclear Physics A}. \BibDash
\newblock 2017. \BibDash
\newblock V. 967. \BibDash
\newblock P.~35--42. \BibDash
\newblock The 26th International Conference on Ultra-relativistic Nucleus-Nucleus Collisions: Quark Matter 2017 URL: \BibUrl{https://www.sciencedirect.com/science/article/pii/S0375947417301197}.

\bibitem{Caines_2004}
\selectlanguageifdefined{english}
\BibEmph{Caines H. et~al.} [STAR Collaboration] An update from STAR at RHIC using strangeness to probe relativistic heavy ion collisions~// \href{http://dx.doi.org/10.1088/0954-3899/30/1/005}{Journal of Physics G: Nuclear and Particle Physics}. \BibDash
\newblock 2003. \BibDash dec. \BibDash
\newblock V.~30, no.~1. \BibDash
\newblock P.~S61. \BibDash
\newblock URL: \BibUrl{https://dx.doi.org/10.1088/0954-3899/30/1/005}.

\bibitem{Shi_2017}
\selectlanguageifdefined{english}
\BibEmph{Shi S., (for~the STAR~collaboration).} Strangeness in STAR experiment at RHIC~// \href{http://dx.doi.org/10.1088/1742-6596/779/1/012008}{Journal of Physics: Conference Series}. \BibDash
\newblock 2017. \BibDash jan. \BibDash
\newblock V. 779, no.~1. \BibDash
\newblock P.~012008. \BibDash
\newblock URL: \BibUrl{https://dx.doi.org/10.1088/1742-6596/779/1/012008}.

\bibitem{Adam-2017}
\selectlanguageifdefined{english}
\BibEmph{{J. Adam, D. Adamová, M. M. Aggarwal et al., }. et~al.} [ALICE Collaboration] Enhanced production of multi-strange hadrons in high-multiplicity proton–proton collisions~// \href{http://dx.doi.org/10.1038/nphys4111}{Nature Physics}. \BibDash
\newblock 2017. \BibDash
\newblock V.~13. \BibDash
\newblock P.~535--539. \BibDash
\newblock URL: \BibUrl{https://doi.org/10.1038/nphys4111}.

\bibitem{PhysRevD.16.2336}
\selectlanguageifdefined{english}
\BibEmph{Okubo S.} Consequences of quark-line (Okubo-Zweig-Iizuka) rule~// \href{http://dx.doi.org/10.1103/PhysRevD.16.2336}{Phys. Rev. D}. \BibDash
\newblock 1977. \BibDash Oct. \BibDash
\newblock V.~16. \BibDash
\newblock P.~2336--2352. \BibDash
\newblock URL: \BibUrl{https://link.aps.org/doi/10.1103/PhysRevD.16.2336}.

\bibitem{PhysRevLett.54.1122}
\selectlanguageifdefined{english}
\BibEmph{Shor A.} $\ensuremath{\varphi}$-Meson Production as a Probe of the Quark-Gluon Plasma~// \href{http://dx.doi.org/10.1103/PhysRevLett.54.1122}{Phys. Rev. Lett.} \BibDash
\newblock 1985. \BibDash Mar. \BibDash
\newblock V.~54. \BibDash
\newblock P.~1122--1125. \BibDash
\newblock URL: \BibUrl{https://link.aps.org/doi/10.1103/PhysRevLett.54.1122}.

\bibitem{Nasim-2015}
\selectlanguageifdefined{english}
\BibEmph{{Md. Nasim, Vipul Bairathi, Mukesh Kumar Sharma, Bedangadas Mohanty and Anju Bhasin}.} A Review on phi-Meson Production in Heavy-Ion Collision Meson Production in Heavy-Ion Collision~// \href{http://dx.doi.org/10.1155/2015/197930}{Advances in High Energy Physics}. \BibDash
\newblock 2015. \BibDash
\newblock V. 2015. \BibDash
\newblock P.~1--16. \BibDash
\newblock URL: \BibUrl{https://doi.org/10.1155/2015/197930}.

\bibitem{PhysRevD.27.140}
\selectlanguageifdefined{english}
\BibEmph{Bjorken J.D.} Highly relativistic nucleus-nucleus collisions: The central rapidity region~// \href{http://dx.doi.org/10.1103/PhysRevD.27.140}{Phys. Rev. D}. \BibDash
\newblock 1983. \BibDash Jan. \BibDash
\newblock V.~27. \BibDash
\newblock P.~140--151. \BibDash
\newblock URL: \BibUrl{https://link.aps.org/doi/10.1103/PhysRevD.27.140}.

\bibitem{PhysRevC.78.044907}
\selectlanguageifdefined{english}
\BibEmph{Alt C. et~al.} [NA49 Collaboration] Energy dependence of $\ensuremath{\phi}$ meson production in central $\mathrm{Pb}+\mathrm{Pb}$ collisions at $\sqrt{{s}_{\mathit{NN}}}=6$ to 17 GeV~// \href{http://dx.doi.org/10.1103/PhysRevC.78.044907}{Phys. Rev. C}. \BibDash
\newblock 2008. \BibDash Oct. \BibDash
\newblock V.~78. \BibDash
\newblock P.~044907. \BibDash
\newblock URL: \BibUrl{https://link.aps.org/doi/10.1103/PhysRevC.78.044907}.

\bibitem{PhysRevLett.87.052301}
\selectlanguageifdefined{english}
\BibEmph{Adcox K. et~al.} [PHENIX Collaboration] Measurement of the Midrapidity Transverse Energy Distribution from $\sqrt{{s}_{\mathrm{NN}}}\phantom{\rule{0ex}{0ex}}=\phantom{\rule{0ex}{0ex}}130\mathrm{GeV}$ $Au+Au$ Collisions at RHIC~// \href{http://dx.doi.org/10.1103/PhysRevLett.87.052301}{Phys. Rev. Lett.} \BibDash
\newblock 2001. \BibDash Jul. \BibDash
\newblock V.~87. \BibDash
\newblock P.~052301. \BibDash
\newblock URL: \BibUrl{https://link.aps.org/doi/10.1103/PhysRevLett.87.052301}.

\bibitem{PhysRevC.88.044910}
\selectlanguageifdefined{english}
\BibEmph{Abelev B. et~al.} [ALICE Collaboration] Centrality dependence of $\ensuremath{\pi}$, $K$, and $p$ production in Pb-Pb collisions at $\sqrt{{s}_{NN}}=2.76$ TeV~// \href{http://dx.doi.org/10.1103/PhysRevC.88.044910}{Phys. Rev. C}. \BibDash
\newblock 2013. \BibDash Oct. \BibDash
\newblock V.~88. \BibDash
\newblock P.~044910. \BibDash
\newblock URL: \BibUrl{https://link.aps.org/doi/10.1103/PhysRevC.88.044910}.

\bibitem{2023137730}
\selectlanguageifdefined{english}
\BibEmph{Acharya S. et~al.} [ALICE Collaboration] System-size dependence of the charged-particle pseudorapidity density at sNN=5.02TeV for pp, pPb, and PbPb collisions~// \href{http://dx.doi.org/https://doi.org/10.1016/j.physletb.2023.137730}{Physics Letters B}. \BibDash
\newblock 2023. \BibDash
\newblock V. 845. \BibDash
\newblock P.~137730. \BibDash
\newblock URL: \BibUrl{https://www.sciencedirect.com/science/article/pii/S0370269323000643}.

\bibitem{PhysRevC.94.034903}
\selectlanguageifdefined{english}
\BibEmph{Adam J. et~al.} [ALICE Collaboration] Measurement of transverse energy at midrapidity in Pb-Pb collisions at $\sqrt{{s}_{\mathit{NN}}}=2.76$ TeV~// \href{http://dx.doi.org/10.1103/PhysRevC.94.034903}{Phys. Rev. C}. \BibDash
\newblock 2016. \BibDash Sep. \BibDash
\newblock V.~94. \BibDash
\newblock P.~034903. \BibDash
\newblock URL: \BibUrl{https://link.aps.org/doi/10.1103/PhysRevC.94.034903}.

\bibitem{PhysRevC.101.044907}
\selectlanguageifdefined{english}
\BibEmph{Acharya S. et~al.} [ALICE Collaboration] Production of charged pions, kaons, and (anti-)protons in Pb-Pb and inelastic $pp$ collisions at $\sqrt{{s}_{NN}}=5.02$ TeV~// \href{http://dx.doi.org/10.1103/PhysRevC.101.044907}{Phys. Rev. C}. \BibDash
\newblock 2020. \BibDash Apr. \BibDash
\newblock V. 101. \BibDash
\newblock P.~044907. \BibDash
\newblock URL: \BibUrl{https://link.aps.org/doi/10.1103/PhysRevC.101.044907}.

\bibitem{Litvinenko-2007}
\selectlanguageifdefined{english}
\BibEmph{Litvinenko A.G.} Some Results Obtained at the Relativistic Heavy Ion Collider~// \href{http://dx.doi.org/10.1134/S1063779607020037}{Physics of Particles and Nuclei}. \BibDash
\newblock 2007. \BibDash
\newblock V.~38. \BibDash
\newblock P.~204–231. \BibDash
\newblock URL: \BibUrl{https://link.springer.com/article/10.1134/S1063779607020037}.

\bibitem{PhysRevLett.116.222302}
\selectlanguageifdefined{english}
\BibEmph{Adam J. et~al.} [ALICE Collaboration] Centrality Dependence of the Charged-Particle Multiplicity Density at Midrapidity in Pb-Pb Collisions at $\sqrt{{s}_{NN}}=5.02\text{ }\text{ }\mathrm{TeV}$~// \href{http://dx.doi.org/10.1103/PhysRevLett.116.222302}{Phys. Rev. Lett.} \BibDash
\newblock 2016. \BibDash Jun. \BibDash
\newblock V. 116. \BibDash
\newblock P.~222302. \BibDash
\newblock URL: \BibUrl{https://link.aps.org/doi/10.1103/PhysRevLett.116.222302}.

\bibitem{Bratkovskaya-2023}
\selectlanguageifdefined{english}
\BibEmph{{Song, Taesoo}., {Aichelin, Joerg}., {Bratkovskaya, Elena}.} In-medium effects on hidden strangeness production in heavy-ion collisions~// \href{http://dx.doi.org/10.1051/epjconf/202327603001}{EPJ Web Conf.} \BibDash
\newblock 2023. \BibDash
\newblock V. 276. \BibDash
\newblock P.~03001. \BibDash
\newblock URL: \BibUrl{https://doi.org/10.1051/epjconf/202327603001}.

\bibitem{PhysRevC.106.034907}
\selectlanguageifdefined{english}
\BibEmph{Acharya S. et~al.} [ALICE Collaboration Collaboration] Production of ${K}^{*}{(892)}^{0}$ and $\ensuremath{\phi}(1020)$ in $pp \mathrm{and} \mathrm{Pb}\text{\ensuremath{-}}\mathrm{Pb}$ collisions at $\sqrt{{s}_{NN}}=5.02 \mathrm{TeV}$~// \href{http://dx.doi.org/10.1103/PhysRevC.106.034907}{Phys. Rev. C}. \BibDash
\newblock 2022. \BibDash Sep. \BibDash
\newblock V. 106. \BibDash
\newblock P.~034907. \BibDash
\newblock URL: \BibUrl{https://link.aps.org/doi/10.1103/PhysRevC.106.034907}.

\bibitem{Stachel_2014}
\selectlanguageifdefined{english}
\BibEmph{Stachel J., Andronic A., Braun-Munzinger P., Redlich K.} Confronting LHC data with the statistical hadronization model~// \href{http://dx.doi.org/10.1088/1742-6596/509/1/012019}{Journal of Physics: Conference Series}. \BibDash
\newblock 2014. \BibDash may. \BibDash
\newblock V. 509, no.~1. \BibDash
\newblock P.~012019. \BibDash
\newblock URL: \BibUrl{https://dx.doi.org/10.1088/1742-6596/509/1/012019}.

\end{thebibliography}

\end{document}